\documentclass{JHEP3}
\usepackage{amssymb}
\usepackage{amsfonts}
\usepackage{mathrsfs}

\newcommand{\e}{\begin{eqnarray}}
\newcommand{\ee}{\end{eqnarray}}

\newcommand{\CN}{{\cal N}}
\newcommand{\CL}{{\cal L}}

\newcommand{\DA}{{\dot \alpha}}
\newcommand{\DB}{{\dot \beta}}
\newcommand{\DC}{{\dot \gamma}}

\def\a{\alpha}
\def\b{\beta}
\def\d{\delta}
\newcommand{\ep}{\epsilon}
\newcommand{\g}{\gamma}
\newcommand{\p}{\psi}
\newcommand{\s}{\sigma}
\def\t{\tau}
\newcommand{\vp}{\varepsilon}

\newcommand{\bp}{\bar{\psi}}

\title{Covariantly Constant Curvature Tensors and $D=3, \CN=4, 5, 8$ Chern-Simons Matter
Theories}

\author{Fa-Min Chen \\
Department of Physics, and State Key Laboratory of Nuclear Physics and Technology,
Peking University, Beijing 100871, China \\
E-mail:
\email{famin.chen@gmail.com}}

\abstract{We construct some examples of $D=3, \CN=4$ GW theory and $\CN=5$ superconformal Chern-Simons matter theory by using the covariantly constant curvature of a quaternionic-Kahler manifold to construct the symplectic 3-algebra in the theories. Comparing with the previous theories, the $\CN=4, 5$ theories constructed in this way possess a local $Sp(2n)$ symmetry and a diffeomorphism symmetry associated with the quaternionic-Kahler manifold. We also construct a generalized $\CN=8$ BLG theory by utilizing the dual curvature operator of a maximally symmetric space of dimension 4 to construct the Nambu 3-algebra. Comparing with the previous $\CN=8$ BLG theory, the theory has a diffeomorphism invariance and a local $SO(4)$ invariance associated with the symmetric space.}


\usepackage{bbm}
\usepackage{amsmath}
\usepackage{slashed}

\begin{document}

\section{Introduction and Summary} \label{SecIntro}

In the last three years, the extended (${\cal N}\geq 4$) supersymmetric
Chern-Simons-matter (CSM) theories in 3D have been constructed by using both ordinary Lie algebras and 3-algebras \cite{Bagger} $\sim$ \cite{Chen:pku1}. In particular, symlectic 3-algebra provides a unified framework for constructing all $\CN\geq 4$ CSM theories \cite{ChenWu3}. Using superalgebra to realize the 3-algebra, one can recover all known examples of the extended $\CN\geq 4$ CSM theories \cite{ChenWu1, Chen2, ChenWu3} and construct several new classes of $\CN=4$ theories as well \cite{ChenWu4}.

A 3-algebra is a triple system. Since a covariantly constant curvature tensor also defines a triple system, it is natural to ask whether it can be used to construct the 3-algebras in the extended CSM theories. In this paper, we demonstrate that at least some special curvature tensor can be used to construct the structure constants of the 3-algebra. Specifically, we use the covariantly constant curvature tensor of a manifold admitting a quaternion structure to construct the symplectic 3-algebra in the $\CN=4$ GW theory and $\CN=5$ theory; the symmetry generated by the curvature tensor is partially gauged, and the resulting gauge group is $Sp(2n)$. Comparing to the original $\CN=4, 5$ theories \cite{GaWi,Hosomichi:2008jb}, the theories constructed in this way have a \emph{local $Sp(2n)$ symmetry} and a \emph{diffeomorphism symmetry} related to the (quaternionic-Kahler) manifold.

We demonstrate that the dual curvature tensor of a $4D$ (internal) manifold also defines a triple system, providing that the curvature tensor is covariantly constant. Furthermore, if the dual curvature tensor is totally antisymmetric, we can use the triple system constructed by the dual curvature operator to realize the Nambu 3-algebra in the $\CN=8$ BLG theory. The $\CN=8$ BLG theory constructed in this way is a generalization of the previous theory in Ref. \cite{Bagger, Gustavsson}, in that it has a \emph{diffeomorphism invariance} and a \emph{local $SO(4)$ symmetry} associated with the $4D$ internal space. The gauge group generated by the dual curvature 3-algebra is still $SO(4)$. It would be nice to analyze this generalized $\CN=8$ BLG theory further.

The paper is organized as follows.  In Section \ref{SecN4N5}, we briefly review the symplectic 3-algebra, and utilize the covariantly constant curvature tensor of a quaternionic-Kahler manifold to construct the symplectic 3-algebra in the $\CN=4$ GW theory  and $\CN=5$ theory \cite{ChenWu3}. In Section \ref{secDCN8}, we use the dual curvature tensor of a maximally symmetric $4D$ space to construct the Nambu 3-algebra in the $\CN=8$ theory. In Appendix \ref{SecRew}, we briefly review the $\CN=4, 5, 8$ CSM theories. Our conventions are summarized in Appendix \ref{secConven}.

\section{Curvature Tensor and Symplectic 3-Algebra}\label{SecN4N5}
\subsection{A Review of Symplectic 3-Algebra}\label{Sec3alg}
In this section, we will review the symplectic 3-algebra \cite{ChenWu1, Chen2}. A symplectic 3-algebra is a complex vector space,
equipped with the 3-bracket
\e
[T_a, T_b; T_c]=f_{abc}{}^dT_d,
\ee
where $T_a$ $(a=1,...,2L)$ is a set of basis generators. We assume that the structure constants are symmetric in the first two indices, i.e.
\e\label{symin2}
f_{abc}{}^d=f_{bac}{}^d.
\ee
The structure constants are required to satisfy the fundamental identity
\e\label{FIF}
f_{abe}{}^gf_{gfcd}+f_{abf}{}^gf_{egcd}-f_{efd}{}^gf_{abcg}-f_{efc}{}^gf_{abdg}=0.
\ee
The transformation of a 3-algebra valued field $X=X^aT_a$ is defined as
\e\label{GlbTran}
\d_{\tilde\Lambda}X^d=\Lambda^{ab}f_{abc}{}^d,
\ee
where $\Lambda^{ab}$ is a set of parameters, satisfying the reality condition
\e\label{reality}
\Lambda^*_{ba}=\Lambda^{ab}=\omega^{ac}\omega^{bd}\Lambda_{cd}.
\ee
To define a symplectic 3-algebra, we require the transform (\ref{GlbTran}) to preserve both the anti-symmetric form $\omega(X,Y)=\omega_{ab}X^aY^b$ and the Hermitian form $h(X,Y)=X^{*a}Y^b$ simultaneously:
\begin{equation}\label{2forms}
\delta_{\tilde\Lambda}\omega(X,Y)=\delta_{\tilde\Lambda}h(X,Y)=0.
\end{equation}
Together with (\ref{symin2}), (\ref{reality}), and (\ref{lconstraint}) below, Eqs. (\ref{2forms}) imply that the structure constants satisfy the symmetry conditions
\e\label{symcndF}
f_{abcd}=f_{bacd}=f_{abdc}=f_{cdab},
\ee
and obey the reality condition
\e\label{rcond}
f^*_{abcd}=f^{abcd}=\omega^{ae}\omega^{bf}\omega^{cg}\omega^{dh}f_{efgh}.
\ee
We have used the invariant antisymmetric tensor $\omega_{ab}$ to lower a 3-algebra index, i.e. $f_{abcd}\equiv \omega_{de}f_{abc}{}^e$. The inverse of $\omega_{ab}$ is denoted as $\omega^{bc}$, satisfying $\omega_{ab}\omega^{bc}=\d_a{}^c$.
Also, to close the $\CN=4,5$ superalgebras, the structure constants must satisfy the linear constraint equation
\e\label{lconstraint}
f_{(abc)d}=0.
\ee
\subsection{Curvature Tensor and Structure Constants of 3-Algebra}\label{secquater}
In this section, we will demonstrate that the covariantly constant curvature tensor of a quaternionic-Kahler manifold can be used to construct the structure constants of the symplectic 3-algebra. Let $(M,g)$ be a $4n$-dimensional manifold, which will be called an internal space. Assume that the metric $g$ is non-degenerate and positive definite. Suppose that the curvature tensor is covariantly constant, i.e.
$
\nabla_IR_{JKLM}=0,
$
with the index $I$ running over
$1,\cdots,4n$. Then the integrability condition
$
[\nabla_I,\nabla_J]R_{KLMN}=0
$
gives
\begin{equation}\label{fndmt}
R^O{}_{KIJ}R_{OLMN}+R^O{}_{LIJ}R_{KOMN}+R^O{}_{MIJ}R_{KLON}+R^O{}_{NIJ}R_{KLMO}=0.
\end{equation}
On the other hand, it is well known that the curvature operator maps three vectors into one vector, that is,
\e\label{cvtop}
R(e_I,e_J)e_K=R_{IJK}{}^Le_L,
\ee
where $e_I$ is a set of basis vectors satisfying \begin{equation}\label{inner0}g(e_I,e_J)=g_{IJ}.\end{equation} Eqs. (\ref{cvtop}) and (\ref{fndmt}) actually define a triple system: using the curvature operator to construct the 3-bracket
\e\label{3bracket}
[e_I,e_J;e_K]\equiv R(e_I,e_J)e_K=R_{IJK}{}^Le_L,
\ee
we see that Eq. (\ref{fndmt}) is equivalent the equation
\begin{equation}\label{FI}
[e_I,e_J; [e_M,e_N;e_K]]=[[e_I,e_J;e_M],e_N;
e_K]+[e_M,[e_I,e_J;e_N]; e_K]+[e_M,e_N; [e_I,e_J;e_K]],
\end{equation}
which plays the role of fundamental identity (FI). We call the Lie triple system defined by (\ref{3bracket}), (\ref{inner0}) and (\ref{FI}) a \emph{curvature 3-algebra}. The curvature 3-algebra can generate an $SO(4n)$ symmetry; the corresponding symmetry group is of course the holonomy group. Writing $R_{IJKL}$ as $(R_{IJ})_{KL}$, we can think  of that $(R_{IJ})$ are a set of
matrices,\footnote{Here $(R_{IJ})$ is \emph{not} the Ricci
tensor $R_{IJ}$.} with $(R_{IJ})_{KL}$ the matrix
elements. Then the matrices $(R_{IJ})$ are indeed a set of $SO(4n)$ generators, since they preserve the symmetric and nondegenerate inner product $g_{KL}$ in the sense that $[\nabla_I,\nabla_J]g_{KL}=R_{IJK}{}^Mg_{ML}+R_{IJL}{}^Mg_{KM}=0$, i.e. the matric elements $(R_{IJ})_{KL}$ are antisymmetric in $KL$. The structure constants of the algebra can be read off from (\ref{fndmt}).

Assume that the manifold admits the quaternion structure or the triplet of complex structures
\e\label{quaternion}\label{quaternion}
(J^i)_I{}^J=-ie_I^{aA}(\s^i)_A{}^B e^J_{aB},
\ee
where $(\s^i)_A{}^B $ ($i=1, 2,3;$  $A=1, 2$) are the pauli matrices. The vielbein $e_I^{aA}$ satisfies
\e
e_I^{aA}e_{JaA}=g_{IJ},\quad e_I^{aA}e^{IbB}=\omega^{ab}\ep^{AB},
\ee
where $e^{IbB}=g^{IJ}e_{J}^{bB}$. Here $\ep^{AB}$ is the antisymmetric tensor of $Sp(2)\cong SU(2)$, and the antisymmetric tensor $\omega^{ab}$ will be identified as the symplectic form of $Sp(2n)$. We denote the inverse of $\ep^{AB}$ as $\ep_{BC}$: $\ep^{AB}\ep_{BC}=\d^A_C$. The inverse of $\omega^{ab}$ is $\omega_{bc}$ satisfying $\omega^{ab}\omega_{bc}=\d^a_c$. Since $g_{IJ}$ is real, the vielbein must obey the reality condition $e_{IaA}=\ep_{AB}\omega_{ab}e_I^{bB}$. The quaternion algebra reads $J^iJ^j=\ep^{ijk}J^k-\d^{ij}$. The triplet of complex structures, vielbein and antisymmetric tensors must be covariant constants,
\e\label{constants}
\nabla_I(J^i)_J{}^K=\nabla_I e_J^{aA}=\nabla_I \ep^{AB}=\nabla_I \omega^{ab}=0.
\ee
The integrability condition
\e\label{dcmBR}
[\nabla_I,\nabla_J]e_K^{aA}=R_{IJK}{}^L e_L^{aA}+R_{IJ}{}^a{}_be_K^{bA}+R_{IJ}{}^A{}_Be_K^{aB}=0
\ee
suggests that the curvature tensor $R_{IJK}{}^L$ can be decomposed into two parts\footnote{For a general discussion of the curvature of quaternionic-Kahler manifolds, see Ref. \cite{Bergshoeff2}.}:
\e\label{dcmR}
e^I_{aA}e^J_{bB}e^K_{cC}e^L_{dD}R_{IJKL}=R_{aA,bB,cC,dD}=\omega_{ab}\omega_{cd}R_{ABCD}+
\ep_{AB}\ep_{CD}R_{abcd}.
\ee
 The symmetry properties of $R_{IJKL}$
($
 R_{IJKL}=-R_{JIKL}=-R_{IJLK}=R_{KLIJ}
$)
  imply that $R_{abcd}$ and $R_{ABCD}$ obey the symmetry conditions
\e\label{symcnd}
&&R_{abcd}=R_{bacd}=R_{abdc}=R_{cdab},\\
\label{symcnd2}&&R_{ABCD}=R_{BACD}=R_{ABDC}=R_{CDAB}.
\ee
The integrability condition
\e
[\nabla_I,\nabla_J]\omega^{ab}=R_{IJ}{}^a{}_c\omega^{cb}+R_{IJ}{}^b{}_c\omega^{ac}=0
\ee
implies that the matrix $R_{IJ}{}^a{}_c$ (for fixed $I$ and $J$) is an $Sp(2n)$ matrix. Similarly, for fixed $I$ and $J$, the matrix $R_{IJ}{}^A{}_B$ is a generator of the Lie algebra of $Sp(2)$. Later we will see, only the symmetry generated by $R_{bd}{}^a{}_c$ will be gauged, meaning that we will gauge \emph{part} of the full symmetry generated by $R_{IJK}{}^L$. By Eqs. (\ref{constants}) and (\ref{quaternion}), we learn that the pauli matrices must be covariantly constant as well, i.e. $\nabla_{aA}(\s^{i})_{EF}=0$. Defining \e(\t_{CD})_{EF}\equiv \s^i_{CD}\s^i_{EF}=\ep_{CE}\ep_{DF}+\ep_{CF}\ep_{DE},\ee
the integrability condition $\omega^{ba}[\nabla_{aA}, \nabla_{bB}](\t_{CD})_{EF}=0$ gives
\begin{equation}\label{pauli}
R_{ABC}{}^G(\t_{GD})_{EF}+R_{ABD}{}^G(\t_{CG})_{EF}+R_{ABE}{}^G(\t_{CD})_{GF}+R_{ABF}{}^G(\t_{CD})_{EG}=0.
\end{equation}

In accordance with the decomposition (\ref{dcmR}), Eq. (\ref{fndmt}) is decomposed into two equations
\e\label{FIR}
&&R_{abe}{}^gR_{gfcd}+R_{abf}{}^gR_{egcd}-R_{efd}{}^gR_{abcg}-R_{efc}{}^gR_{abdg}=0,\\
&&R_{ABE}{}^GR_{GFCD}+R_{ABF}{}^GR_{EGCD}-R_{EFD}{}^GR_{ABCG}-R_{EFC}{}^GR_{ABDG}=0.\label{FIR2}
\ee
It can be see that (\ref{symcnd}) and (\ref{FIR}) take exactly the same forms as that of (\ref{symcndF}) and (\ref{FIF}), respectively. However, if we want to identify the structure constants $f_{abcd}$ with $R_{abcd}$, we must make sure that $R_{abcd}$ also obeys the linear constraint equation (\ref{lconstraint}) and satisfies the reality condition (\ref{rcond}). We will see that at least in some special case, these two requirements can be fulfilled. To see this, let us consider the algebraic property of the Riemann curvature tensor
\e\label{cycR}
R_{aA,bB,cC,dD}+R_{bB,cC,aA,dD}+R_{cC,aA,bB,dD}=0.
\ee
Using the decomposition (\ref{dcmR}), Eq. (\ref{cycR}) can be converted into
\e\label{cyc}
&&R_{abcd}\ep_{AB}\ep_{CD}+R_{bcad}\ep_{BC}\ep_{AD}+R_{cabd}\ep_{CA}\ep_{BD}\nonumber\\
&&+R_{ABCD}\omega_{ab}\omega_{cd}+R_{BCAD}\omega_{bc}\omega_{ad}+R_{CABD}\omega_{ca}\omega_{bd}=0.
\ee
Let us solve for $R_{ABCD}$ first; comparing (\ref{FIR2}) with (\ref{pauli}), we find an obvious solution to these two equations:
\e\label{sp2m}
R_{ABCD}=k(\t_{AB})_{CD}=k(\ep_{AC}\ep_{BD}+\ep_{AD}\ep_{BC}),
\ee
where $k$ is proportional to the (constant) curvature scalar $R=g^{IJ}R_{IJ}$. It can be seen that right hand side of (\ref{sp2m}) satisfies the symmetry conditions (\ref{symcnd2}) and Eq. (\ref{FIR2}). Substituting equation (\ref{sp2m}) above into (\ref{cyc}), we obtain
\e
&&[R_{abcd}-R_{bcad}-k(\omega_{bc}\omega_{ad}-2\omega_{ca}\omega_{bd}+\omega_{ab}\omega_{cd})]
\ep_{AB}\ep_{CD}\nonumber\\&&+
[R_{cabd}-R_{bcad}+k(\omega_{bc}\omega_{ad}-2\omega_{ab}\omega_{cd}+\omega_{ca}\omega_{bd})]
\ep_{CA}\ep_{BD}=0,
\ee
where we have used the identity $\ep_{AB}\ep_{CD}=\ep_{AC}\ep_{BD}-\ep_{BC}\ep_{AD}$. We observe that if the first line vanishes, then the second line vanishes automatically, and vice versa. We therefore need only to consider the equation
\begin{equation}
R_{abcd}-R_{bcad}-k(\omega_{bc}\omega_{ad}-2\omega_{ca}\omega_{bd}-\omega_{ab}\omega_{cd})=0.
\end{equation}
Under the condition $R_{(abc)d}=0$, the solution is given by
\begin{equation}\label{sp2nm}
R_{abcd}=k(\omega_{ac}\omega_{bd}+\omega_{ad}\omega_{bc}),
\end{equation}
which is nothing but an $Sp(2n)$ matrix (for fixed $a$ and $b$). Now it is straightforward to check that (\ref{sp2nm}) obeys the linear constraint equation (\ref{lconstraint}) and satisfies the reality condition (\ref{rcond}): namely, $R_{(abc)d}=0$ and $R^*_{abcd}=R^{abcd}=\omega^{ae}\omega^{bf}\omega^{cg}\omega^{dh}R_{efgh}$. Hence $R_{abcd}$ can be used to construct the structure constants of the symplectic 3-algebra. Substituting (\ref{sp2m}) and  (\ref{sp2nm}) into (\ref{dcmR}) determines $k=\frac{R}{8n(n+2)}$.
Also by (\ref{sp2m}) and (\ref{sp2nm}), we learn that our solution $R_{aA,bB,cC,dD}$ is consisted of entirely by covariantly constant quantities such as $\omega_{ab}$ and $\ep_{AB}$, so it must be also a covariantly constant tensor, i.e. $\nabla_{I}R_{aA,bB,cC,dD}=0$. Setting $f_{abcd}=R_{abcd}$ and substituting (\ref{sp2nm}) into Eq. (\ref{LN4GW}) (Eq. (\ref{GeneN5Lagran})) gives the $\CN=4$ GW (\CN=5) theory with $Sp(2n)$ gauge group. 

It can be seen that the $\CN=4$ action constructed here has the symmetries associated with the quaternionic-Kahler manifold:
\begin{itemize}
\item Diffeomorphism invariance\footnote{The indices $\a=1,2$ and $\dot\a=1,2$ below denote the bifundamental representation of the R-symmetry group $SU(2)\times SU(2)$ (see Section \ref{secN4}).}:
\begin{equation}
R^\prime_{abcd}(q^\prime)=R_{abcd}(q),\quad Z^{\prime a}_\a(q^\prime)=Z^a_\a(q), \quad \psi^{\prime a}_{\dot\a}(q^\prime)=\psi^a_{\dot\a}(q),\quad A^{\prime ab}_\mu(q^\prime)=A^{ab}_\mu(q)
\end{equation}
with $q^J$ a set of local coordinates, and $q^I\rightarrow q^{\prime I}$ an arbitrary coordinate transformation.
\item Local $Sp(2n)$ symmetry:
\e\label{local}
&&\hat Z_\a^a(q)=L^a{}_b(q)Z_\a^b(q),\quad \hat \psi_{\dot\a}^a(q)=L^a{}_b(q)\psi_{\dot\a}^b(q),\quad \hat A^{ab}_\mu(q)=L^{a}{}_c(q)L^b{}_d(q)A^{cd}_\mu(q),\nonumber\\
&&\hat f_{abcd}(q)=\hat R_{abcd}(q)=L_a{}^{e}(q)L_b{}^{f}(q)L_c{}^{g}(q)L_d{}^{h}(q)R_{efgh}(q)=R_{abcd}(q),
\ee
where $L_a{}^{e}(q)=\omega_{ac}\omega^{ed}L^{c}{}_d(q)$, and $L^a{}_b(q)$ satisfies
\e\label{sp2ntrans}
L^{c}{}_a(q)L^{d}{}_b(q)\omega_{cd}=\omega_{ab}.
\ee
In the last equation of (\ref{local}), we have used (\ref{sp2nm}) and (\ref{sp2ntrans}).
\end{itemize}
Similarly, the $\CN=5$ theory also possesses the diffeomorphism symmetry and the local $Sp(2n)$ symmetry associated with the internal space.


However, we emphasis that the $\CN=4$ GW theory constructed here is \emph{not} a conventional nonlinear sigma model like the one in Ref. \cite{{GaWi}}: in our construction, the scalar fields $Z_\a^a$  are a set of \emph{complex vectors} of the \emph{quaternionic-Kahler manifold}, while in the original $\CN=4$ GW non-linear sigma model, the scalar fields are a set of \emph{local coordinates} of the target space being a $4n$-dimensional \emph{hyper-Kahler manifold}. Also, the gauge symmetry of the $\CN=4$ GW theory constructed here is generated by the curvature 3-algebra or the \emph{holonomy algebra} of the internal space, while in the $\CN=4$ GW nonlinear sigma model,
the gauge symmetry is generated by the
\emph{Killing vectors} of the target space.

\section{Dual Curvature Tensor and Generalized $\CN=8$ BLG Theory}\label{secDCN8}
In this section, we will demonstrate that the Nambu 3-algebra can be realized by utilizing the dual curvature operator of a $4D$ maximally symmetric space. We call this symmetric space an internal space. A generalized $\CN=8$ BLG theory possessing a diffeomorphism invariance and a local $SO(4)$ symmetry related to the internal space can be constructed by virtue of the dual curvature tensor.

In $4D$, the dual curvature tensor is defined as
\e\label{ddual}
\tilde R_{abcd}=\frac{1}{2}\sqrt g\vp_{abef}R^{ef}{}_{cd},
\ee
where $g={\rm det}(g_{ab})$, and $\sqrt g\vp_{abef}$ is the totally antisymmetric tensor\footnote{Here $a=1,\cdots, 4$ is a tangent vector index, \emph{not} an $Sp(2n)$ fundamental index of the last section. We hope this will not cause any confusion.}. (We assume that the metric is nondegenerate and positive definite.) If the curvature tensor $R_{efcd}$ satisfies $\nabla_gR_{efcd}=0$, we must have $\nabla_g\tilde R_{abcd}=0$ on account of that $\sqrt g\vp_{abef}$ is always a covariant constant. To prove that $\nabla_g(\sqrt g\vp_{abef})=0$, we introduce a set of vielbein fields $e^a_i$ ($i=1,\cdots,4$) satisfying $\d^{ij}e^a_ie^b_j=g^{ab}$ and $g_{ab}e^a_ie^b_j=\d_{ij}$. Now the totally antisymmetric tensor can be converted into a constant $\vp_{ijkl}=e^a_ie^b_je^c_ke^d_l\sqrt g\vp_{abcd}$ (our convention is that $\vp^{ijkl}=\d^{im}\d^{jn}\d^{ko}\d^{lp}\vp_{mnop}=\vp_{ijkl}$), and its covariant derivative is given by
\begin{equation}\label{Dvp}
\nabla_m\vp_{ijkl}=e^a_m(\partial_a\vp_{ijkl}-\omega_a{}^n{}_i\vp_{njkl}-\omega_a{}^n{}_j\vp_{inkl}
-\omega_a{}^n{}_k\vp_{ijnl}-\omega_a{}^n{}_l\vp_{ijkn}).
\end{equation}
The spin connection can be written as $\omega_a{}^n{}_i=\frac{1}{2}\omega_a^{op}(\s_{op}){}^n{}_i,$
with $(\tau_{op}){}^n{}_i=\d^n_o\d_{pi}-\d_{oi}\d^n_p$ the $SO(4)$ matrices. Since $\partial_a\vp_{ijkl}=0$, the right hand side of (\ref{Dvp}) becomes
\begin{equation}\label{Dvp2}
-\frac{1}{2}e^a_m\omega_a^{op}[(\tau_{op}){}^n{}_i\vp_{njkl}+(\tau_{op}){}^n{}_j\vp_{inkl}
+(\tau_{op}){}^n{}_k\vp_{ijnl}+(\tau_{op}){}^n{}_l\vp_{ijkn}].
\end{equation}
The quantity in the bracket is nothing but the variation of $\vp_{ijkl}$ under the transformation generated by $(\tau_{op})$; it vanishes due to the fact that $\vp_{ijkl}$ is $SO(4)$-invariant. Alternatively, one can write the first term in the bracket of (\ref{Dvp2}) as
\e\label{1st}
(\tau_{op})_{ni}\vp_{njkl}=(\frac{1}{2}\vp_{msop}\vp^{ms}{}_{ni})\vp_{njkl}.
\ee
Substituting $\vp_{opms}\vp^{njkl}=\d_o^{[n}\d_p^j\d_m^k\d_s^{l]}$ into the right hand side of (\ref{1st}) proves that (\ref{Dvp2}) is zero. This completes the proof that $\nabla_m\vp_{ijkl}=0$, meaning that the tensor $\sqrt g\vp_{abef}$ is a covariant constant. Assuming that $\nabla_aR_{cdef}=0$, and multiplying the integrability condition $[\nabla^g,\nabla^h]\tilde R_{cdef}=0$ by $\frac{1}{2}\sqrt g\vp_{abgh}$, we obtain
\e\label{FI4NS}
\tilde R_{abe}{}^g\tilde R_{gfcd}+\tilde R_{abf}{}^g\tilde R_{egcd}-\tilde R_{efd}{}^g\tilde R_{abcg}-\tilde R_{efc}{}^g\tilde R_{abdg}=0.
\ee
On the other hand, we can construct a 3-bracket in terms of the dual curvature operator:
\e\label{dulop}
\{e_a, e_b, e_c\}\equiv \frac{1}{2}\sqrt g\vp_{abef}R(e^e,e^f)e_c=\tilde R_{abc}{}^de_d,
\ee
with $e_a$ a set of basis vectors satisfying
\begin{equation}\label{inner}
g(e_a,e_b)=g_{ab}.
\end{equation}
We now see that taking account of the inner product (\ref{inner}), Eq. (\ref{FI4NS}) is equivalent to the fundamental identity
\begin{equation}\label{FI4N}
\{e_a,e_b, \{e_c,e_d,e_e\}\}=\{\{e_a,e_b,e_c\},e_d,
e_e\}+\{e_c,\{e_a,e_b,e_d\}, e_e\}+\{e_c,e_d, \{e_a,e_b,e_e\}\}.
\end{equation}
We call the triple system defined by Eqs. (\ref{dulop}), (\ref{inner}) and (\ref{FI4N}) a \emph{dual curvature 3-algebra}. In Section \ref{secquater}, we have demonstrated that the curvature tensor of the $4n$-dimensional manifold can generate an $SO(4n)$ symmetry. Based on the same reason, the dual curvature 3-algebra can generate an $SO(4)$ symmetry.

If $\tilde R_{abcd}$ is completely antisymmetric in all indices, the dual curvature 3-algebra is an obvious realization of the Nambu 3-algebra\footnote{If the inner product is Lorentzian in the sense that $g^{ab}e_a^ie_b^j=\eta^{ij}$, then it is a realization of the Lorentzian 3-algebra. But we do not consider this case in the current paper.}. We now assume that $\tilde R$ is totally antisymmetric. Since in $4D$ the totally antisymmetric tensor is essentially unique, we must have
$
\tilde R_{abcd}=k\sqrt g\vp_{abcd}.
$
Using Eq. (\ref{ddual}), one can determine that
\e\label{slvr}
R_{efcd}=\frac{R}{12}(g_{ec}g_{fd}-g_{ed}g_{fc}).
\ee
Namely, the manifold is maximally symmetric and $k$ is given by $k=\frac{R}{12}$, with $R$ the curvature scalar, which must be a constant. Therefore our final result is
\e\label{trf}
f_{abcd}=\tilde R_{abcd}=\frac{R}{12}\sqrt g\vp_{abcd}.
\ee
$\tilde R_{abcd}$ being totally antisymmetric is a necessary condition for closing the $\CN=8$ superalgebra. We now present an alternative derivation of (\ref{trf}). Since $\tilde R_{abcd}=-\tilde R_{bacd}=-\tilde R_{abdc}$, $\tilde R_{abcd}$ will be totally antisymmetric if $\tilde R_{abcd}=-\tilde R_{acbd}$. Multiplying both sides of the equation by $\frac{1}{2\sqrt g}\vp_{ef}{}^{ab}$ and using (\ref{ddual}), we obtain
\e
R_{efcd}=(\frac{1}{2\sqrt g}\vp_{ef}{}^{ab})(-\frac{1}{2}\sqrt g\vp_{ac}{}^{gh}R_{ghbd}).
\ee
A short calculation gives
\e\label{msr}
R_{efcd}+\frac{1}{3}(g_{fc}R_{ed}-g_{ec}R_{fd})=0.
\ee
In order that $R_{efcd}=-R_{efdc}$, we must require that $g_{fc}R_{ed}-g_{ec}R_{fd}=-(g_{fd}R_{ec}-g_{ed}R_{fc})$. Multiplying both sides by $g^{cf}$ determines the Ricci tensor $R_{ed}$ uniquely: $R_{ed}=\frac{R}{4}g_{ed}$. Substituting it into (\ref{msr}) gives (\ref{slvr}).
Combining (\ref{slvr}) and (\ref{ddual}), we obtain (\ref{trf}) again.

We see that the curvature tensor (\ref{slvr}) indeed obeys the crucial equation $\nabla_bR_{efcd}=0$. The dual curvature tensor satisfies Eq. (\ref{FI4NS}), and has the desired  symmetry properties as well. So (\ref{dulop}) and (\ref{FI4N}), as well as the inner product (\ref{inner}), are indeed a realization of the Nambu 3-algebra. In this realization, we must use the metric $g_{ab}$ and its inverse $g^{bc}$ to lower and raise indices, respectively. Plugging (\ref{trf}) into Eq. (\ref{N6Lagrangian}) and (\ref{N6susy}) gives the $\CN=8$ BLG theory with $SO(4)$ gauge group. The matter fields are in the vector representation of $SO(4)$. The $\CN=8$, $SO(4)$ theory has been
conjectured to be the dual gauge theory of two M2-branes \cite{ABJM}.

Comparing with the original $\CN=8$ BLG theory in Ref. \cite{Bagger, Gustavsson}, our theory has a diffeomorphism invariance and a local $SO(4)$ symmetry related to the $4D$ (internal) symmetric space. Specifically, the theory is invariant under the transformations
\e
&&Z_{a}^{\prime A}(\sigma^\prime)=\frac{\partial\s^{b}}{\partial\s^{\prime a}}Z_b^A(\s),\quad \psi^{\prime }_{Aa}(\sigma^\prime)=\frac{\partial\s^{b}}{\partial\s^{\prime a}}\psi_{Ab}(\s),
\nonumber\\&&  A^\prime_\mu{}^{a}{}_{b}(\s^\prime)=\frac{\partial\s^{\prime a}}{\partial\s^{c}}\frac{\partial\s^d}{\partial\s^{\prime b}}A_\mu{}^{c}{}_{d}(\s),
\quad g^\prime_{ab}(\s^\prime)=\frac{\partial\s^c}{\partial\s^{\prime a}}\frac{\partial\s^d}{\partial\s^{\prime b}}g_{cd}(\s),
\nonumber
\\&&
f^\prime_{abcd}(\s^\prime)=\tilde R^\prime_{abcd}(\s^\prime)=\frac{\partial\s^e}{\partial\s^{\prime a}}
\frac{\partial\s^f}{\partial\s^{\prime b}}\frac{\partial\s^g}{\partial\s^{\prime c}}
\frac{\partial\s^h}{\partial\s^{\prime d}}\tilde R_{efgh}(\s)=\sqrt{g^\prime}\vp_{abcd},
\ee
where $\s^a$ and $\s^{\prime a}$ are two sets of coordinates of the $4D$ internal space. The $\CN=8$ action is also invariant under the local $SO(4)$ transformations:
\e\nonumber
&&\hat Z^A_i(\s)=L_i{}^j(\s)Z^A_j(\s),\quad \hat\psi_{Ai}(\s)=L_i{}^j(\s)\psi_{Aj}(\s),\quad \hat A_\mu{}^i{}_j(\s)=L^i{}_k(\s)L_j{}^l(\s)A_\mu{}^k{}_l(\s)\\
&&L^k{}_i(\s)L^l{}_j(\s)\d_{kl}=\d_{ij},\quad \hat f_{ijkl}(\s)=\hat{\tilde{R}}_{ijkl}(\s)=\vp_{ijkl}=\tilde R_{ijkl}(\s),
\ee
where $Z^A_i=e^a_i Z^A_a$, $\psi_{Ai}=e^a_i\psi_{Aa}$ and $A_\mu{}^i{}_j=e^i_be^b_jA_\mu{}^a{}_b$, and $L_i{}^j(\s)=\d_{ik}\d^{jl}L^k{}_l(\s)$.
Hence our theory is sort of generalized $\CN=8$ BLG theory.

\section{Acknowledgement}

We are grateful to Yong-Shi Wu for useful discussion; we thank the referee for useful comments.

\appendix
\section{A Review the $\CN=4, 5, 8$ Theories.}\label{SecRew}
In this section, we review the $\CN=4, 5, 8$ CSM theories.
\subsection{$\CN=4$ GW Theory}\label{secN4}
The $\CN=4$ GW theory was first constructed in Ref. \cite{GaWi}, using an ordinary
Lie algebra approach. In Ref. \cite{ChenWu3}, the $\CN=4$ GW theory was constructed
in terms of the symplectic 3-algebra. (The symplectic 3-algebra is reviewed in Sec. \ref{Sec3alg}.) The action reads
\e\label{LN4GW}\CL&=&\frac{1}{2}(-D_\mu\bar{Z}^\a_aD^\mu
Z^a_\a+i\bp^\DA_a\g^\mu
D_\mu\p^a_\DA)-\frac{i}{2}f_{acbd}Z^a_\a Z^{\a b}\p^c_\DB\p^{\DB
d}\nonumber\\&&+\frac{1}{2}\epsilon^{\mu\nu\lambda}(f_{abcd}A_\mu^{ab}\partial_\nu
A_\lambda^{cd}+\frac{2}{3}f_{abc}{}^gf_{gdef}A_\mu^{ab}A_\nu^{cd}A_\lambda^{ef})\nonumber\\
&&+\frac{1}{12}f_{abcg}f^g{}_{def}Z^{\a a}Z^b_\b Z^{\b(c}Z^{d)}_\g Z^{\g e}Z^f_\a.
\ee
Here $\a=1,2$ and $\dot\a=1,2$ are the undotted  and dotted indices of the $SU(2)\times SU(2)$ R-symmetry group,
respectively; $a=1,\cdots, 2n$ a \emph{symplectic} 3-algebra index. The covariant derivative is defined as
\e
 D_\mu Z^\a_d =
\partial_\mu Z^\a_d -\tilde A_\mu{}^c{}_dZ^\a_c ,\quad
\tilde A_\mu{}^c{}_d=A^{ab}_\mu f_{ab}{}^c{}_d.
\ee
The matter fields obey the natural reality conditions $
\bar Z^\a_a=\omega_{ab}\ep^{\a\b}Z^b_\b,$ and $\bar
\p^\DA_a=\omega_{ab}\ep^{\DA\DB}Z^b_\DB$.
Here $\ep^{\a\b}$ and $\epsilon^{\DA\DB}$ are invariant antisymmetric tensors of the R-symmetry group $SU(2)\times SU(2)$, satisfying $\ep^{\a\b}\ep_{\b\g}=\d^\a_\g$ and $\ep^{\DA\DB}\ep_{\DB\DC}=\d^\DA_\DC$.
The supersymmetry transformations are given by
\e \label{SUSY4GW}&&\delta Z^a_\a=i\ep_\a{}^\DA\p^a_\DA,\nonumber\\
&&\delta\p^a_\DA=-\g^\mu D_\mu
Z^a_\b\ep^\dag_\DA{}^\b-\frac{1}{3}f^a{}_{bcd}Z^b_\b
Z^{\b c}Z^d_\g\ep^\dag_\DA{}^\g,\nonumber\\
&&\delta\tilde A_\mu{}^c{}_d=i\ep^{\a\DB}\g_\mu\p^b_\DB
Z^a_\a f_{ab}{}^c{}_d,
\ee
where the parameter satisfies the reality condition
\e
\ep^{\dag}{}_{\DA}{}^{\b}=-\epsilon^{\b\g}\epsilon_{\DA\DB}\ep{}_{\g}{}^{\DB}.
\ee

\subsection{$\CN=5$  Theory}\label{SecN5}
The $\CN=5$ action reads \cite{Chen2}
\begin{eqnarray}\label{GeneN5Lagran}\nonumber
{\cal L}&=&\frac{1}{2}(-D_\mu
\bar Z_a^\a D^{\mu}Z^a_\a+i\bar{\psi}_a^\a D_\mu\gamma^\mu\psi^a_\a)\nonumber\\
&&-\frac{i}{2}\omega^{\a\b}\omega^{\g\d}\omega_{de}f_{abc}{}^e(Z^a_\a Z^c_\b\bar{\psi}^b_\g\psi^d_\d-
2Z^a_\a Z^c_\d\bar{\psi}^b_\g\psi^d_\b)\nonumber\\
&&+\frac{1}{2}\epsilon^{\mu\nu\lambda}(\omega_{de}f_{abc}{}^eA_\mu^{ab}\partial_\nu
A_\lambda^{cd}+\frac{2}{3}\omega_{fh}f_{abc}{}^gf_{gde}{}^hA_\mu^{ab}A_\nu^{cd}A_\lambda^{ef})\\
&&-\frac{1}{60}(2f_{abc}{}^gf_{gdf}{}^e-9f_{cda}{}^gf_{gfb}{}^e+2f_{abd}{}^gf_{gcf}{}^e)Z^f_\a
Z^{\a a}Z^b_\b Z^{\b c}Z^d_\g Z^{\g}_e.\nonumber
\end{eqnarray}
Here $\a=1,\cdots,4$ is a fundamental index of the $Sp(4)$ R-symmetry group
\footnote{In the $\CN=4$ theory (see Appendix \ref{secN4}), $\a=1,2$
transforms in the fundamental representation of one factor of the $SU(2)\times SU(2)$ R-symmetry group.
We hope this will not cause any confusion.};
$a=1,\cdots, 2n$ a \emph{symplectic} 3-algebra index. The covariant derivative is defined as $D_\mu Z^\a_d =
\partial_\mu Z^\a_d -\tilde A_\mu{}^c{}_dZ^\a_c ,$ where $
\tilde A_\mu{}^c{}_d=A^{ab}_\mu f_{ab}{}^c{}_d$. The matter fields obey the reality conditions
\begin{eqnarray}\label{RealCondi}
Z_{\a}^{*a}=\omega^{\a\b}\omega_{ab}Z^b_\b ,\quad
\psi_{\a}^{*a}=\omega^{\a\b}\omega_{ab}\psi^b_\b.
\end{eqnarray}
Here $\omega_{\a\b}$ is the invariant antisymmetric tensor of $Sp(4)$, satisfying $\omega_{\a\b}\omega^{\b\g}=\d^\g_\a$. The supersymmetry transformations are given by
\begin{eqnarray}\label{GeneSusyTransLaw}\nonumber
\delta Z^a_\a&=&i\bar{\epsilon}_\a{}^\b\psi^a_\b\nonumber\\
\delta\psi^a_\a&=&\gamma^{\mu}D_\mu Z^a_\b\epsilon^\b{}_\a
+\frac{1}{3}f_{cdb}{}^a\omega^{\b\g}Z^b_\b Z^c_\g Z^d_\d\epsilon^\d{}_\a
-\frac{2}{3}f_{cdb}{}^a\omega^{\b\d}Z^b_\g Z^c_\d Z^d_\a\epsilon^\g{}_\b
\nonumber\\
\delta \tilde{A}_\mu{}^c{}_d &=&
i\bar{\epsilon}^{\a\b}\gamma_\mu\psi^b_\b Z^a_\a f_{abd}{}^c,
\end{eqnarray}
where the parameter $\epsilon^{\a\b}$ is antisymmetric in $\a\b$,
satisfying
\begin{eqnarray}\label{SusyPara5}\nonumber
\omega_{\a\b}\epsilon^{\a\b}=0 ,\quad
\epsilon^{*}_{\a\b}=\omega^{\a\g}\omega^{\b\d}\epsilon_{\g\d}.
\end{eqnarray}

\subsection{$\CN=8$ BLG Theory}
Here we will follow the convention of Ref. \cite{Chen:pku1}. The action is given by
\begin{eqnarray}\label{N6Lagrangian}
\nonumber {\cal L} &=& -D_\mu
\bar{Z}_A^aD^\mu Z^A_a - i\bar\psi^{Aa}\gamma^\mu D_\mu\psi_{Aa}\\
\nonumber && -if^{ab}{}_{cd}\bar\psi^{Ad} \psi_{Aa}
Z^B_b\bar{Z}_B^c+2if^{ab}{}_{cd}\bar\psi^{Ad}
\psi_{Ba}Z^B_b\bar{Z}_A^c\\ \nonumber
&&-\frac{i}{2}\varepsilon_{ABCD}f^{ab}{}_{cd}\bar\psi^{Ac}
\psi^{Bd}Z^C_aZ^D_b -\frac{i}{2}\varepsilon^{ABCD}f^{cd}{}_{ab}
\bar\psi_{Ac}\psi_{Bd}\bar{Z}_C^a\bar{Z}_D^b \\
&&+\frac{1}{2}\varepsilon^{\mu\nu\lambda}
(f^{ab}{}_{cd}A_{\mu}{}^c{}_b\partial_\nu A_{\lambda}{}^d{}_a
+\frac{2}{3}f^{ac}{}_{dg}f^{ge}{}_{fb}
A_{\mu}{}^b{}_aA_{\nu}{}^d{}_c A_{\lambda}{}^f{}_e)\\ \nonumber &&
-\frac{2}{3}(f^{ab}{}_{cd}f^{ed}{}_{fg}
-\frac{1}{2}f^{eb}{}_{cd}f^{ad}{}_{fg})\bar{Z}_A^c Z^A_e\bar{Z}_B^f
Z^B_a\bar{Z}_D^g Z^D_b.
\end{eqnarray}
Here $A=1,\cdots,4$ is a fundamental index of the $SU(4)$ R-symmetry group\footnote{In Section \ref{secquater}, $A=1,2$ denotes the $Sp(2)$ index of the curvature tensor of the
quarternionic-Kahler manifold; in this section, $A=1,\cdots, 4$ refers to the fundamental index of the $SU(4)$ R-symmetry of the BLG theory. We hope this will not cause any confusion.}. And $a=1,\cdots, L$ is a \emph{Hermitian} 3-algebra index. The covariant derivative is defined as $D_\mu Z^A_b =
\partial_\mu Z^A_b -\tilde A_\mu{}^a{}_bZ^A_a ,$ where $\tilde{A}_\mu{}^a{}_b= A_{\mu}{}^d{}_cf^{ac}{}_{db}$.
The SUSY transformation law reads
\begin{eqnarray}\label{N6susy}
\nonumber  \delta Z^A_d &=& -i\bar\epsilon^{AB}\psi_{Bd} \\
 \nonumber
\delta \psi_{Bd} &=& \gamma^\mu D_\mu Z^A_d\epsilon_{AB} +
  f^{ab}{}_{cd}Z^C_aZ^A_b \bar{Z}_{C}^{c} \epsilon_{AB}+f^{ab}{}_{cd}
  Z^C_a Z^D_{b} \bar{Z}_{B}^{c}\epsilon_{CD} \\
 \delta \tilde A_\mu{}^c{}_d &=&
-i\bar\epsilon_{AB}\gamma_\mu Z^A_a\psi^{Bb} f^{ca}{}_{bd} +
i\bar\epsilon^{AB}\gamma_\mu \bar{Z}_{A}^{a}\psi_{Bb}f^{cb}{}_{ad}.
\end{eqnarray}
Here the SUSY transformation parameters $\epsilon_{AB}$ satisfy
\begin{equation}
\epsilon_{AB}=-\epsilon_{BA},\quad
\epsilon^*_{AB}=\epsilon^{AB}
=\frac{1}{2}\varepsilon^{ABCD}\epsilon_{CD}.
\end{equation}
In the action and the supersymmetry transformation law, only $SU(4)$ R-symmetry is manifest. However, in Ref. \cite{Bagger10}, it was demonstrated explicitly that theory actually has an $\CN=8$ R-symmetry, if the structure constants $f^{abcd}$ are totally antisymmetric or the Hermitian 3-algebra becomes the Nambu 3-algebra.

\section{Conventions}\label{secConven}

In $1+2$ dimensions, the gamma matrices are defined as
\begin{equation}
(\gamma_{\mu})_{\alpha}{}^\gamma(\gamma_{\nu})_{\gamma}{}^\beta+
(\gamma_{\nu})_{\alpha}{}^\gamma(\gamma_{\mu})_{\gamma}{}^\beta=
2\eta_{\mu\nu}\delta_{\alpha}{}^\beta.
\end{equation} For the metric we
use the $(-,+,+)$ convention.
We also define the totally antisymmetric tensor
$\varepsilon^{\mu\nu\lambda}=-\varepsilon_{\mu\nu\lambda}$. So
$\varepsilon_{\mu\nu\lambda}\varepsilon^{\rho\nu\lambda} =
-2\delta_\mu{}^\rho$. We raise and lower spinor indices with an
antisymmetric matrix
$\epsilon_{\alpha\beta}=-\epsilon^{\alpha\beta}$, with
$\epsilon_{12}=-1$. For example,
$\psi^\alpha=\epsilon^{\alpha\beta}\psi_\beta$ and
$\gamma^\mu_{\alpha\beta}=\epsilon_{\beta\gamma}(\gamma^\mu)_\alpha{}^\gamma
$, where $\psi_\beta$ is a Majorana spinor.
We use the following spinor summation convention:
$
\psi\chi=\psi^\alpha\chi_\alpha,$ $
\psi\gamma_\mu\chi=\psi^\alpha(\gamma_{\mu})_{\alpha}{}^\beta\chi_\beta,
$
where $\psi$ and $\chi$ are anti-commuting Majorana spinors.

\end{document}